\documentclass[final,5p,times,twocolumn,number]{elsarticle}

\usepackage{amssymb}
\usepackage{lipsum}
\usepackage{graphicx}
\usepackage{subcaption}
\usepackage{url}
\usepackage{amsmath}
\usepackage{float}
\usepackage{placeins}
\usepackage{booktabs}

\usepackage{microtype}

\makeatletter
\def\ps@pprintTitle{%
\let\@oddhead\@empty
\let\@evenhead\@empty
\let\@oddfoot\@empty
\let\@evenfoot\@oddfoot
}
\makeatother

\begin{document}

	\begin{frontmatter}
		
		
		\title{Process-Aware Thickness Analysis in CAD Models using Hybrid Geometric Methods}
		
		\author[first]{Serafeim Baltadouros\textsuperscript{*}}
		\author[first]{Joost R. Duflou}
		
		\address[first]{KU Leuven, Department of Mechanical Engineering, Division of Manufacturing Processes and Systems (MaPS) and Flanders Make@KU Leuven, 3001 Leuven, Belgium
		\par\vspace{6pt}
		\begin{minipage}{\linewidth}
			\raggedright\footnotesize
			$^{*}$Corresponding author. Tel.: +30-694-585-9222.
			\textit{E-mail address:} serafeim.baltadouros@kuleuven.be
			\end{minipage}}

		\begin{abstract}
			Thickness is a critical geometric attribute in Design for Manufacturability (DFM), yet its computational analysis in CAD environments remains largely process-agnostic. Existing approaches rely either on inscribed-sphere or ray-casting methods, each carrying limitations that make them poorly suited as universal solutions across manufacturing processes. This paper presents a process-aware thickness analysis system for parts intended for molding and milling, where the geometric method is selected and its outputs interpreted according to the DFM rules meaningful to each process. For molding, the sphere-based method detects maximum thickness violations and wall non-uniformity, and segments parts into distinct thickness zones. For milling, ray-casting identifies regions of insufficient thickness and detects thin features prone to deflection or failure. Validation on representative parts demonstrates that this hybrid approach surfaces process-specific manufacturability issues, offering designers actionable geometric feedback early in the design cycle.
		\end{abstract}
		
		
	\end{frontmatter}

	
	\section{Introduction}
	
	Design for Manufacturability (DFM) is the practice of evaluating a part's 
	geometry against the constraints imposed by its intended manufacturing 
	process, with the goal of identifying potential defects before production 
	begins \cite{boothroyd1994}. Among the geometric attributes relevant to 
	DFM, wall thickness is one of the most consequential. In molding operations, 
	thickness directly governs material flow, cooling behavior, and the onset of 
	defects such as sink marks and warpage \cite{osswald2001}, while in machining 
	it determines the mechanical rigidity of features during cutting, with 
	insufficiently thin walls prone to deflection, chatter, and failure 
	\cite{boothroyd1994}.
	
	Automated thickness analysis has been an active area of research in 
	CAD and computer-aided engineering (CAE). Two families of methods have 
	emerged as dominant. In the ray-based method, 
	the thickness at a surface point is measured by casting a ray inward along 
	the surface normal and recording the distance to the first opposing surface 
	intersection \cite{beiter1991, cocks1983, sinha2007}. In the sphere-based 
	method, thickness is defined as the diameter of the largest inscribed sphere 
	that fits within the solid and touches the surface point in question 
	\cite{lambourne2005, inui2016}. Each method carries distinct strengths and 
	limitations: the ray method is computationally straightforward but produces 
	ambiguous results when opposing surfaces are not parallel \cite{inui2016}, 
	while the sphere method yields consistent volumetric thickness estimates but 
	tends to underestimate thickness near convex corners and sharp edges 
	\cite{inui2016, inui2024}.
	
	Despite this body of work, existing tools and studies rarely account for 
	the fact that the choice of geometric method and the DFM rules it is used 
	to evaluate are not independent decisions. The criteria that define a 
	manufacturable part differ substantially between processes: a molded part 
	must respect maximum wall thickness limits and maintain uniform wall 
	sections to avoid solidification defects, while a machined part must avoid 
	regions too thin to withstand cutting forces and must not contain tall, 
	slender features vulnerable to deflection. Applying a single method 
	uniformly, or evaluating thickness without reference to process-specific 
	rules, reduces the diagnostic value of the analysis and risks either 
	missing critical issues or flagging irrelevant ones. Automated systems 
	have been developed for standalone process DFM rule check in CAD geometry 
	\cite{malyshev2022}, yet the coupling between method selection and 
	process-specific rule interpretation has not been systematically addressed.
	
	This paper addresses that gap by presenting a process-aware thickness 
	analysis system in which the geometric method is selected and its results 
	interpreted in accordance with the DFM rules of the target process. For 
	molding, the system detects maximum 
	thickness violations, wall non-uniformity, and distinct thickness zones 
	within the part. For milling, it flags regions of 
	insufficient absolute thickness and identifies thin features at risk of 
	failure. The system operates on standard STL and STEP file formats, making 
	it compatible with common CAD workflows.

	\section{State of the Art}
	
	Thickness analysis of three-dimensional CAD models has been an active 
	area of research in both geometric computing and manufacturing engineering. 
	Early approaches to thickness measurement in a manufacturing context were 
	motivated by practical DFM needs: the need to predict sink marks in molded 
	parts \cite{beiter1991}, to assess wall sections in die cast components 
	\cite{cocks1983}, and to support concurrent engineering by providing 
	designers with geometric feedback early in the design cycle. As CAD systems 
	matured, thickness analysis was gradually integrated into commercial 
	platforms. Tools such as GeomCaliper \cite{sinha2007}, adopted as an add-on 
	in CATIA and SolidWorks \cite{inui2016}, provide thickness visualization 
	directly on tessellated CAD models. More recently, automated DFM systems 
	have emerged that combine thickness analysis with feature recognition to 
	flag manufacturability issues for specific processes \cite{malyshev2022}. 
	Across this body of work, two computational methods have established 
	themselves as dominant: the ray-based method and the sphere-based method 
	\cite{sinha2007, inui2016}.
	
	The ray-based method defines thickness at a surface point $p$ as the 
	distance from $p$ to the point $q$ where a ray cast inward along the 
	surface normal first intersects the opposing surface \cite{beiter1991, 
		cocks1983, sinha2007}. For tessellated models, the computation reduces to 
	a ray-triangle intersection problem, which can be accelerated using spatial 
	data structures such as uniform grids or k-d trees \cite{sinha2007}. The 
	per-face independence of the computation makes it naturally parallelisable, 
	and GPU implementations have demonstrated practical performance on 
	high-resolution meshes \cite{inui2016}. The method has been applied in 
	DFM contexts primarily to identify thin wall sections, where insufficient 
	material between two opposing surfaces presents a structural or 
	manufacturing risk \cite{beiter1991, cocks1983}. Sinha \cite{sinha2007} 
	provided a systematic treatment of the ray method within the GeomCaliper 
	system, noting that it is particularly effective at detecting thin regions 
	and measuring nominal wall thickness in parts with regular geometry. 
	Malyshev and Tcherniavski \cite{malyshev2022} demonstrated its use in 
	an automated DFM system for machined parts, combining ray-based thickness 
	queries with accessibility analysis to flag regions that are too thin or 
	too narrow to machine reliably.
	
	The principal limitation of the ray method is its geometric ambiguity 
	when opposing surfaces are not parallel. Since the inward normal at $p$ 
	may not be perpendicular to the opposing surface, the thickness measured 
	at $p$ and the thickness measured at the opposing point $q$ will in 
	general differ, yielding an inconsistent result \cite{inui2016}. On 
	curved or inclined surfaces this inconsistency can be significant. 
	Furthermore, in geometrically complex regions where no directly opposing 
	surface exists along the normal direction, the ray may travel a large 
	distance before intersecting any surface, returning a spuriously large 
	thickness value \cite{sinha2007}. The method is therefore most reliable 
	on parts with nominally planar, parallel opposing walls, and its 
	diagnostic value degrades on freeform or highly curved geometries.
	
	The sphere-based method defines thickness at a surface point $p$ as 
	the diameter of the largest sphere that can be inscribed within the 
	solid and touches $p$ tangentially \cite{lambourne2005, inui2016}. 
	This definition is closely related to the Medial Axis Transform, as 
	the locus of the centres of all maximal inscribed spheres traces the 
	medial axis of the object \cite{lambourne2005}. Lambourne et al. 
	\cite{lambourne2005} presented one of the earliest algorithms for 
	computing inscribed sphere thickness directly on trimmed NURBS surfaces, 
	reformulating the problem as a minimisation over surface points and 
	solving it iteratively using a bounded quasi-Newton method. Their 
	approach avoids the explicit construction of Voronoi diagrams or 
	Delaunay tetrahedra required by earlier medial axis methods, and 
	allows thickness to be evaluated at individual surface points. Inui 
	et al. \cite{inui2016} later proposed the shrinking sphere algorithm 
	for tessellated models, an iterative method that begins with a large 
	candidate sphere touching $p$ and progressively shrinks it by 
	identifying the surface polygon most constraining the sphere radius 
	at each iteration. Accelerated by a hierarchical axis-aligned bounding 
	box tree and parallelised on the GPU using CUDA, the algorithm 
	converges in typically fewer than five iterations and was demonstrated 
	on meshes of nearly two million polygons. The sphere method has been 
	applied in molding DFM contexts to detect regions of material 
	accumulation linked to sink marks and warpage \cite{osswald2001, 
		inui2018}, where it outperforms the ray method by providing consistent 
	volumetric thickness estimates regardless of surface orientation.
	
	The sphere method is not without limitations. Near convex edges and 
	sharp corners, the largest inscribed sphere is geometrically 
	constrained by the proximity of multiple surfaces simultaneously, 
	causing the method to systematically underestimate thickness in these 
	regions \cite{inui2016, inui2024}. This effect is inherent to the 
	inscribed sphere definition and cannot be resolved by increasing mesh 
	resolution. Inui et al. \cite{inui2024} proposed an alternative based 
	on maximum inscribed cubes to partially address this, though the 
	corner underestimation problem remains a known failure mode of 
	sphere-based approaches. Additionally, the iterative nature of 
	inscribed sphere computation carries a higher computational cost per 
	surface point than the ray method, making it less attractive for 
	interactive or large-scale analysis without GPU acceleration 
	\cite{inui2016}.
	
	Beyond the ray and sphere families, a third class of methods 
	represents the solid as a volumetric data structure and computes 
	thickness through operations on this discrete representation. Patil 
	and Ravi \cite{patil2005} proposed a voxel-based system supporting 
	progressive cross-section display, layer removal simulation, and 
	radiographic projection for thickness visualization of complex 
	freeform shapes, achieving voxelisation and analysis of models at 
	resolutions up to one billion voxels. Subburaj et al. 
	\cite{subburaj2006} extended voxel-based approaches with multiple 
	thickness metrics including interior thickness derived from distance 
	fields, exterior thickness based on the object skeleton, and 
	radiographic thickness. Distance-field methods, in which each voxel 
	records its distance to the nearest surface boundary, have similarly 
	been used for thickness and clearance analysis \cite{inui2015}. 
	Despite their ability to handle complex freeform geometries and their 
	flexibility in defining thickness metrics, volumetric methods carry 
	substantial memory requirements for adequate resolution, are sensitive 
	to discretisation error in thin regions, and do not naturally operate 
	on the surface-based representations standard in CAD workflows. For 
	these reasons, volumetric approaches have seen limited adoption in 
	commercial CAD and DFM tools, and are outside the scope of the 
	present work.
	
	Despite the maturity of individual thickness analysis methods, the 
	reviewed literature reveals a consistent gap: the choice of geometric 
	method and the DFM rules applied to its output are treated as 
	independent concerns, with no systematic coupling between method 
	selection and the failure modes of the target manufacturing process. 
	The following section addresses this gap by describing a methodology 
	in which both the thickness estimation method and the rule evaluation 
	criteria are determined by the intended process.

	\section{Methodology}
	
\begin{figure}[t]  
		\centering
		\includegraphics[width=0.5\textwidth]{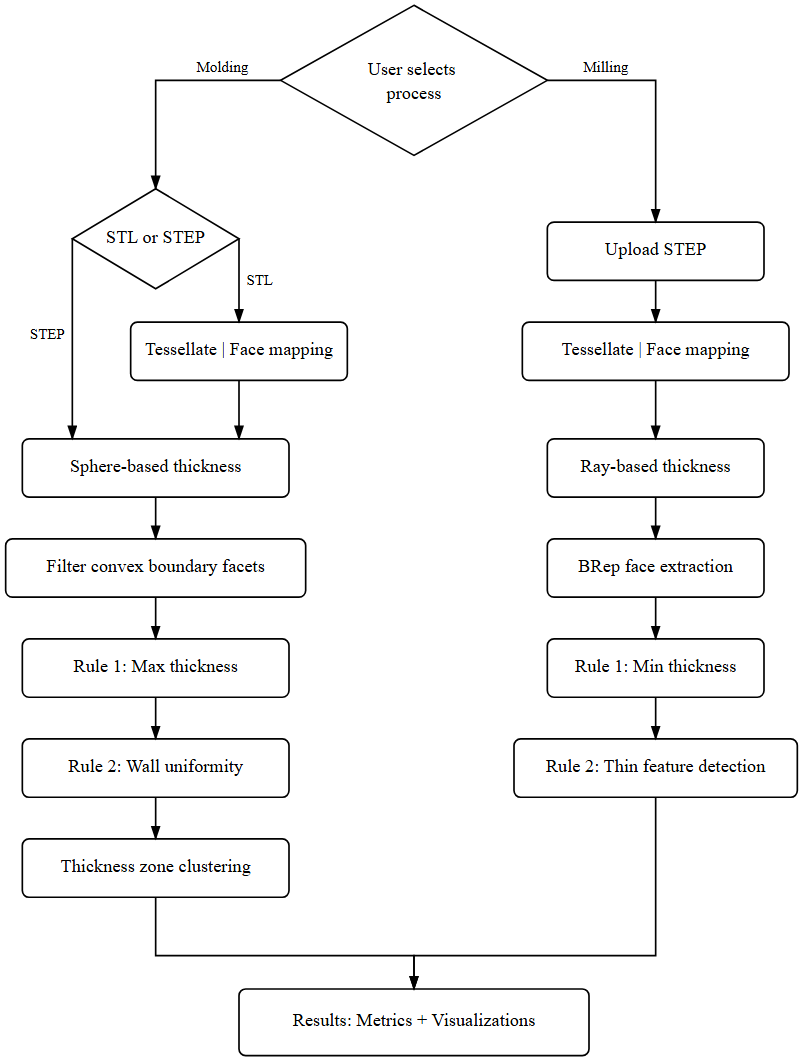}  
		\caption{Pipeline flowchart}
		\label{fig:flowchart}  
	\end{figure}

	The proposed system performs thickness analysis in two distinct branches, 
	each corresponding to a target manufacturing process: molding and milling. 
	The process selection determines both the geometric method applied and the 
	DFM rules evaluated. For molding, the input is a triangulated surface mesh 
	in STL format, and thickness is estimated via an inscribed-sphere approach. 
	For milling, the input is a boundary representation model in STEP format, 
	which is first converted to an STL mesh before ray-based thickness 
	estimation is performed. In both branches, thickness computation is 
	GPU-accelerated using NVIDIA Warp \cite{warp2022}, enabling practical 
	performance on high-resolution meshes. The overall structure of the system 
	is depicted in Figure~\ref{fig:flowchart}.
	
	The selection of thickness estimation method per process is deliberate and 
	grounded in the complementary limitations of each approach. The sphere-based 
	method tends to underestimate thickness near convex edges and corners 
	\cite{inui2016, inui2024}, making it unreliable for minimum thickness 
	assessment. However, it produces consistent and accurate volumetric 
	estimates in regions of material accumulation, making it well suited to 
	detecting maximum thickness violations and non-uniformity in molded parts, 
	where these are the primary DFM concerns. Conversely, the ray-based method 
	can overestimate thickness when the ray does not encounter a directly 
	opposing surface, returning spuriously large values in geometrically complex 
	regions \cite{sinha2007}. However, it accurately captures the 
	through-thickness distance between nominally parallel opposing surfaces, 
	which is precisely the quantity of interest in milling, where minimum 
	absolute thickness and feature slenderness are the critical failure 
	indicators. Each method is therefore applied where its strengths align with 
	the DFM rules of the target process, and its known failure modes fall 
	outside the region of interest.
	
	\subsection{Molding Analysis}
	
	The molding analysis branch accepts an STL file as input. The mesh is 
	loaded and the centroid and outward surface normal of each triangular face 
	are computed. The inward normal, defined as the negation of the outward 
	normal, is used to direct the thickness estimation inward into the solid.
	
\begin{figure}[t]
		\begin{subfigure}{0.24\textwidth}
			\includegraphics[width=\textwidth]{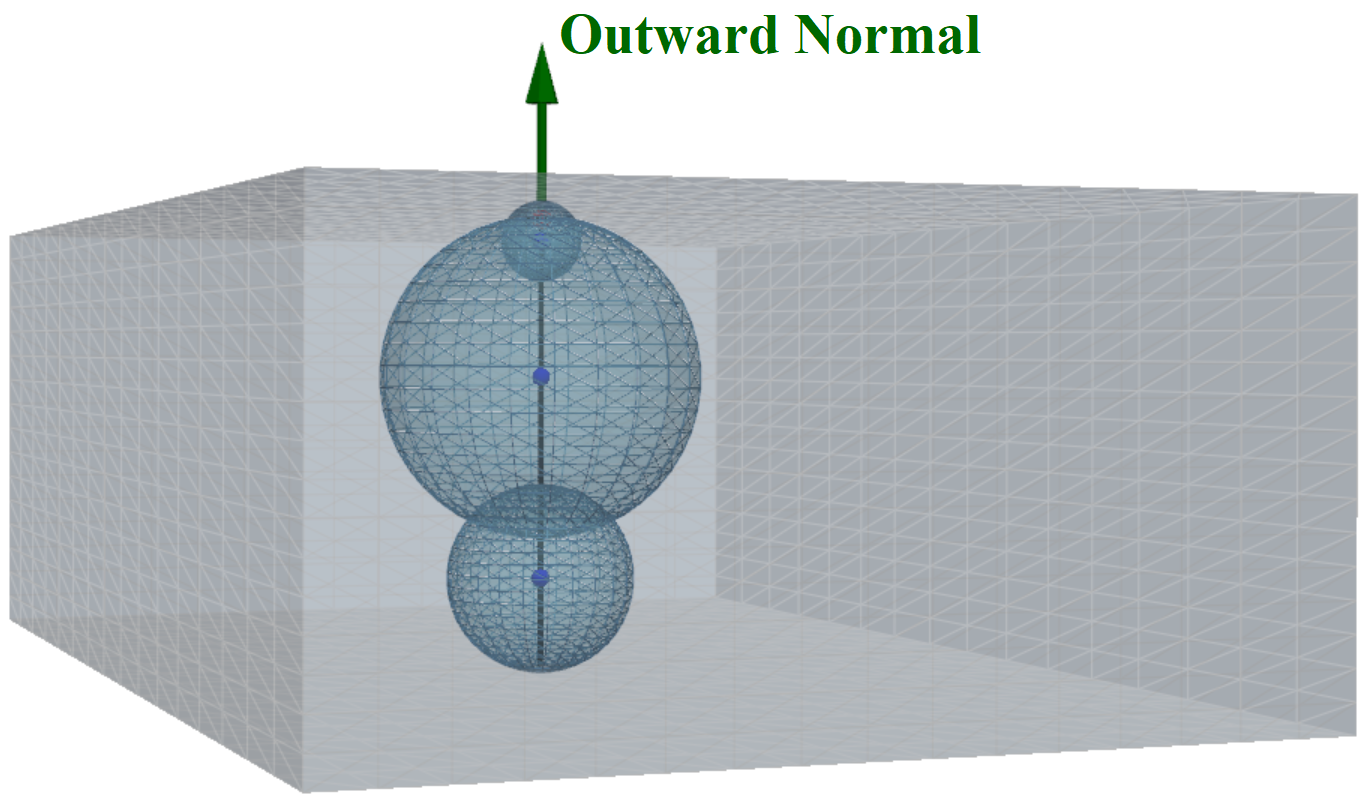}
			\caption{}
		\end{subfigure}
		\hfill
		\begin{subfigure}{0.235\textwidth}
			\includegraphics[width=\textwidth]{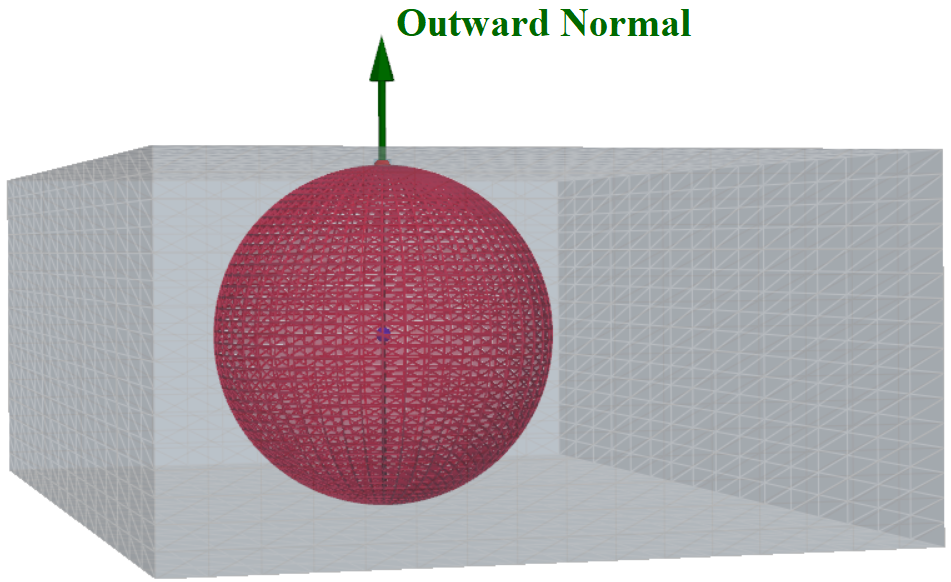}
			\caption{}
		\end{subfigure}
		\caption{Inscribed sphere thickness estimation method  (a) Sample spheres (b) Largest sphere}
		\label{fig:molding_sphere} 
	\end{figure}

	Thickness at each face is estimated using an inscribed-sphere approach 
	\cite{inui2016}. Rather than iteratively growing a sphere, the method 
	approximates the largest inscribed sphere by sampling candidate sphere 
	centers along the inward ray emanating from each face centroid. Specifically, 
	a ray is first cast from each face centroid along the inward normal using 
	GPU-accelerated ray casting, identifying the distance to the first opposing 
	surface intersection. This ray length defines the extent of the solid in 
	that direction. A set of $n$ sample points is then placed at uniform 
	intervals along this ray segment, where $n$ is a user-defined parameter 
	controlling the resolution of the sphere estimation. For each sample point, 
	the distance to the nearest surface point on the mesh is computed on the 
	GPU using a nearest-point query, yielding the radius of the largest sphere 
	centered at that point. The thickness assigned to the face is taken as twice 
	the maximum radius found across all sample points along the ray, 
	corresponding to the diameter of the largest inscribed sphere 
	approximated along that inward path. The process is shown in 
	Figure~\ref{fig:molding_sphere}. Figure~\ref{fig:molding_sphere}a shows 3 sample points/spheres along the inward ray to illustrate how the
	method samples candidate sphere centers. Figure~\ref{fig:molding_sphere}b shows only the single largest inscribed sphere found across all $n$ sample points, i.e. the sphere that
	actually defines the estimated thickness.
	
	Prior to rule evaluation, faces whose estimated thickness falls below an
	automatically computed minimum threshold are excluded. This threshold is
	derived from the mean sphere-based thickness of faces located near the
	bounding box boundary of the part, scaled by a user-defined edge factor.
	At convex edges and corners, the largest inscribed sphere is geometrically
	constrained by the proximity of multiple surfaces, causing the sphere-based
	method to systematically underestimate thickness in these regions
	\cite{inui2016, inui2024}. The boundary face mean provides a local estimate
	of this underestimation, and faces falling below the derived threshold are
	excluded as geometrically unreliable. The remaining faces are considered
	valid for analysis.
	
	Two DFM rules are evaluated on the valid face set. Rule~1 checks whether 
	any face exceeds a user-defined maximum thickness threshold. Faces violating 
	this rule are flagged, and the percentage of violating faces over the total 
	valid set is reported. Thick regions increase material usage, extend cooling 
	time, and are a primary cause of sink marks in molded parts \cite{osswald2001}. 
	Rule~2 checks wall thickness uniformity. For each pair of spatially adjacent 
	faces, identified via a k-d tree on face centroids \cite{inui2016}, the 
	relative thickness change is computed. A pair is flagged as non-uniform if 
	the relative difference exceeds a user-defined tolerance. Abrupt thickness 
	transitions promote residual stress and warpage during solidification 
	\cite{osswald2001}.
	
	In addition to rule evaluation, valid faces are clustered into distinct 
	thickness zones using a bin-merging procedure. Face thickness values are 
	first discretised into bins of fixed width, and adjacent bins whose mean 
	values fall within a merge tolerance are iteratively combined. The merge 
	tolerance is automatically adjusted until the number of resulting zones 
	falls within a user-defined target range. Each zone groups faces of 
	approximately equal thickness, providing the designer with a segmented 
	view of the thickness distribution across the part. Per-zone statistics including mean, 
	minimum, maximum, and standard deviation of thickness, as well as the 
	fraction of faces exceeding the maximum thickness threshold, are reported.
	
	\subsection{Milling Analysis}
	
	The milling analysis branch accepts a STEP file as input. Since thickness 
	computation operates on a triangulated surface, the STEP model is first 
	converted to an STL mesh using the open-source meshing library Gmsh 
	\cite{geuzaine2009}. The target triangle count is user-defined, and mesh 
	sizing parameters are automatically estimated from a coarse initial mesh 
	of the geometry. During meshing, a face map is constructed that records, 
	for each STEP surface, the set of STL triangle indices corresponding to 
	that surface. This mapping is saved as a JSON file alongside the STL and 
	is reused in subsequent analyses of the same part, avoiding redundant 
	recomputation.

\begin{figure}[t]  
		\centering
		\includegraphics[width=0.3\textwidth]{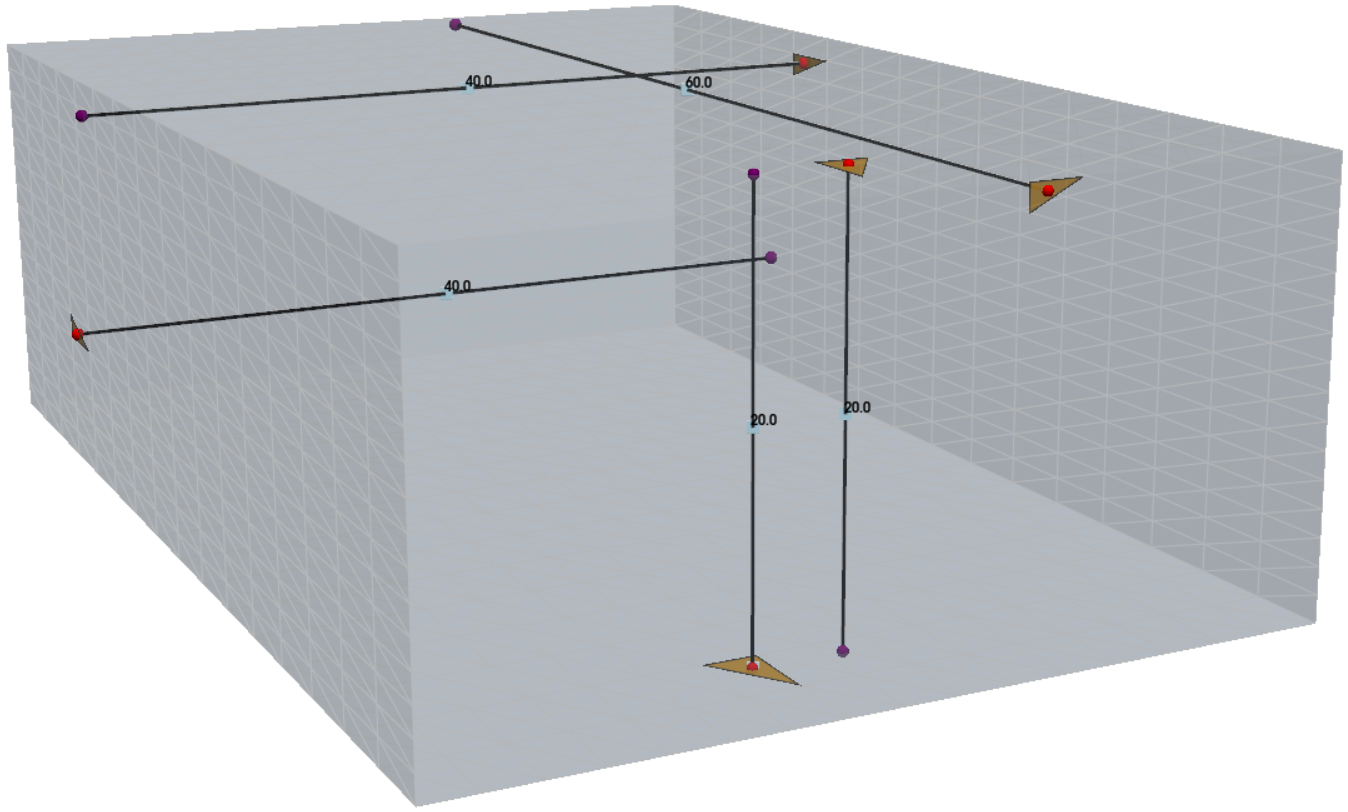}  
		\caption{Ray-cast thickness estimation method}
		\label{fig:millinganalysis}  
	\end{figure}

	Thickness at each triangle is estimated by ray casting. From the centroid 
	of each triangle, a ray is cast along the inward surface normal, as depicted in Figure~\ref{fig:millinganalysis}, using the 
	GPU-accelerated ray casting kernel. The distance to the first opposing 
	surface intersection is recorded as the thickness at that triangle. 
	Triangles for which no intersection is found are excluded from further 
	analysis. As discussed, the ray-based method may return overestimated 
	thickness values in geometrically complex regions; however, since the 
	analysis targets minimum thickness violations and thin feature detection, 
	falsely large values do not affect rule evaluation and are 
	effectively inert in this context.
	
	To support thin feature detection, the STEP boundary representation is 
	parsed using the Open CASCADE Technology (OCCT) kernel \cite{occt}. For 
	each STEP face, surface type, area, bounding box dimensions, and, for 
	planar faces, the surface normal and a point on the plane are extracted. 
	Each STEP face is then assigned a mean thickness computed as the average 
	ray-cast thickness of all STL triangles mapped to it. A height value is 
	also assigned to each face, defined as the largest bounding box dimension 
	orthogonal to the face normal, representing the extent of the face in its 
	dominant direction. The height-to-thickness ratio of each face is then 
	computed as the ratio of this height to its mean thickness, providing a 
	geometric indicator of slenderness.
	
	Two DFM rules are evaluated. Rule~1 flags triangles whose ray-cast 
	thickness falls below user-defined warning and critical thresholds. 
	Regions of insufficient absolute thickness are structurally at risk 
	during machining, as thin walls exhibit reduced stiffness and are prone 
	to deflection under cutting forces \cite{boothroyd1994}. Results are 
	classified per triangle as critical, warning, or acceptable, and 
	aggregate counts and percentages are reported.
	
	Rule~2 detects thin features by identifying pairs of opposing planar 
	STEP faces. Two faces are considered an opposing pair if their surface 
	normals are approximately antiparallel, their mean thicknesses are 
	similar, and the line segment connecting their centroids passes through 
	the interior of the part without intersecting any other surface. Interior 
	passage is verified by casting a ray between the two face centroids and 
	checking for intermediate intersections, and additionally by confirming 
	that a set of uniformly spaced points along the connecting segment all 
	lie within the solid using a six-direction ray parity test on the GPU. 
	For each detected pair, the face separation distance is computed as the 
	projection of the inter-centroid vector onto the shared face normal, and 
	the mean height-to-thickness ratio of the pair is used to assess 
	slenderness. Pairs exceeding user-defined warning and critical 
	height-to-thickness thresholds are flagged accordingly.This rule targets thin-walled or 
	rib-like features that, while individually passing the absolute 
	thickness check, present a geometric configuration susceptible to 
	deflection and chatter during milling \cite{boothroyd1994}.
	
	Both rule results are propagated back to the triangle level for 
	visualization, with each triangle colored according to its worst-case 
	classification across all applicable rules. A thickness heatmap, 
	per-rule diagnostic views, and a thin feature overlay showing detected 
	face pairs with their geometric annotations are made available for 
	inspection.

	\section{Case Study Validation}

	The following case studies apply the proposed analysis system to two real 
	industrial parts, each intended for a different manufacturing process. Both 
	parts have been slightly modified to better showcase the diagnostic 
	capabilities of the system. The molding case study part is sourced from 
	\cite{flandersmake}, while the milling case study part is an openly 
	available CNC test part \cite{grabcad2023}. Analysis parameters are 
	adjusted in each case to demonstrate the system's sensitivity and 
	configurability, though a designer would set these according to their own 
	material specifications and process constraints. The results presented 
	here are reported without interpretation, which is reserved for the 
	discussion section that follows.
	
	\subsection{Molding Case Study}
\begin{figure}[h]  
		\centering
		\includegraphics[width=0.35\textwidth]{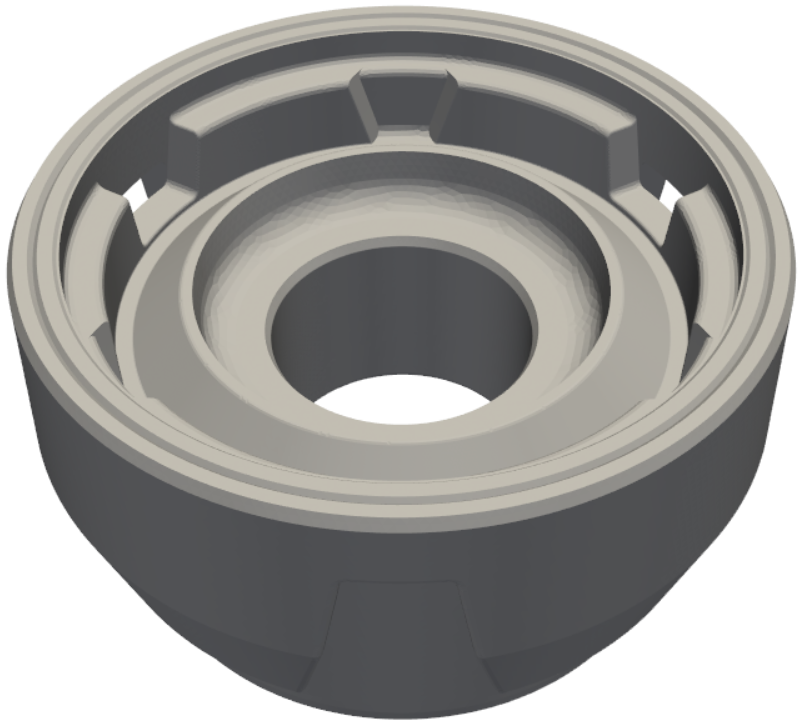}  
		\caption{Molding - Case Study Part}
		\label{fig:moldingpart}  
	\end{figure}

\begin{figure}[h!]  
		\centering
		\includegraphics[width=0.35\textwidth]{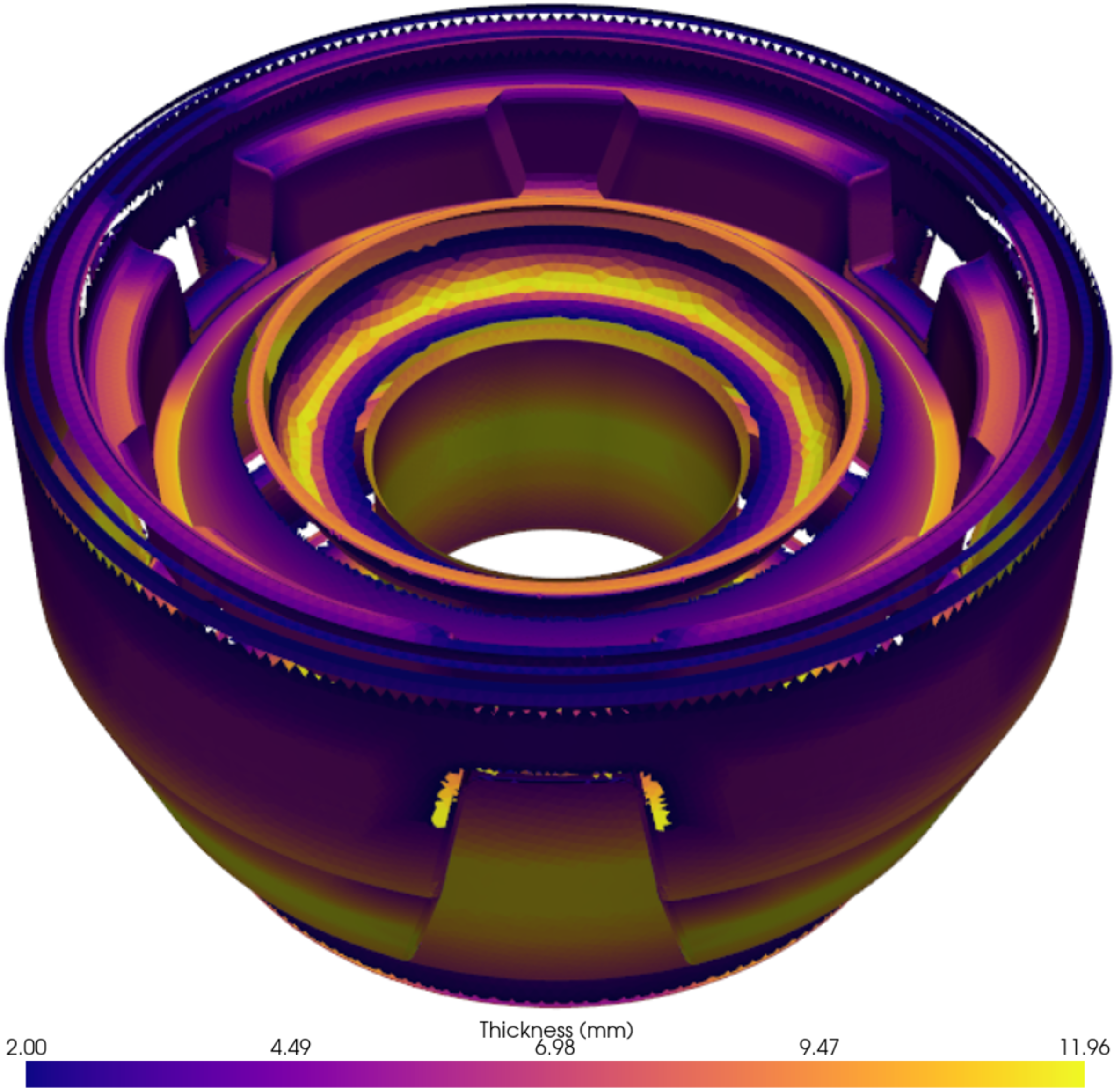}  
		\caption{Sphere-based Thickness Heatmap}
		\label{fig:heatmapmolding}  
	\end{figure}

	The molding case study part is shown in Figure~\ref{fig:moldingpart}. 
	The analysis is performed on the imported STL file of the part, using the 
	parameters listed in Table~\ref{tab:moldingparams}. The resulting 
	sphere-based thickness heatmap is shown in Figure~\ref{fig:heatmapmolding}. 
	According to the summary statistics reported in Table~\ref{tab:moldingstats}, 
	93.8\% of the analyzed facets are valid for analysis, meaning that 6.2\% of 
	the STL facets are below the minimum threshold set by analyzing the 
	thickness values of triangles adjacent to the boundary convex edges of the 
	part, which, by definition of sphere-based analysis, would return false 
	values.
	
	\begin{table}[h]
		\centering
		\begin{tabular}{l c}
			\hline
			\textbf{Parameter} & \textbf{Value} \\
			\hline
			Maximum thickness threshold & 5.0~mm\\
			Sphere sections & 50\\
			Uniformity tolerance & 60\%\\
			Target zone count & 1--4\\
			\hline
		\end{tabular}
		\caption{Analysis parameters for the molding case study.}
		\label{tab:moldingparams}
	\end{table}

	\begin{table}[h]
		\centering
		\begin{tabular}{l c}
			\hline
			\textbf{Metric} & \textbf{Value} \\
			\hline
			Minimum thickness & 1.72~mm\\
			Maximum thickness & 29.99~mm\\
			Mean thickness & 6.63~mm\\
			Standard deviation & 2.68~mm\\
			Valid facets (of total) & 93.8\%\\
			\hline
		\end{tabular}
		\caption{Thickness statistics for the molding case study (valid face set).}
		\label{tab:moldingstats}
	\end{table}

	\begin{table}[h]
		\centering
		\begin{tabular}{l p{5cm} c}
			\hline
			\textbf{Rule} & \textbf{Criterion} & \textbf{Facets flagged} \\
			\hline
			Rule 1 & Exceeds max.\ thickness threshold & 74.7\%\\
			Rule 2 & Non-uniform thickness & 5.9\%\\
			\hline
		\end{tabular}
		\caption{Rule evaluation results for the molding case study.}
		\label{tab:moldingrules}
	\end{table}

	\begin{figure}[h!]  
		\centering
		\includegraphics[width=0.35\textwidth]{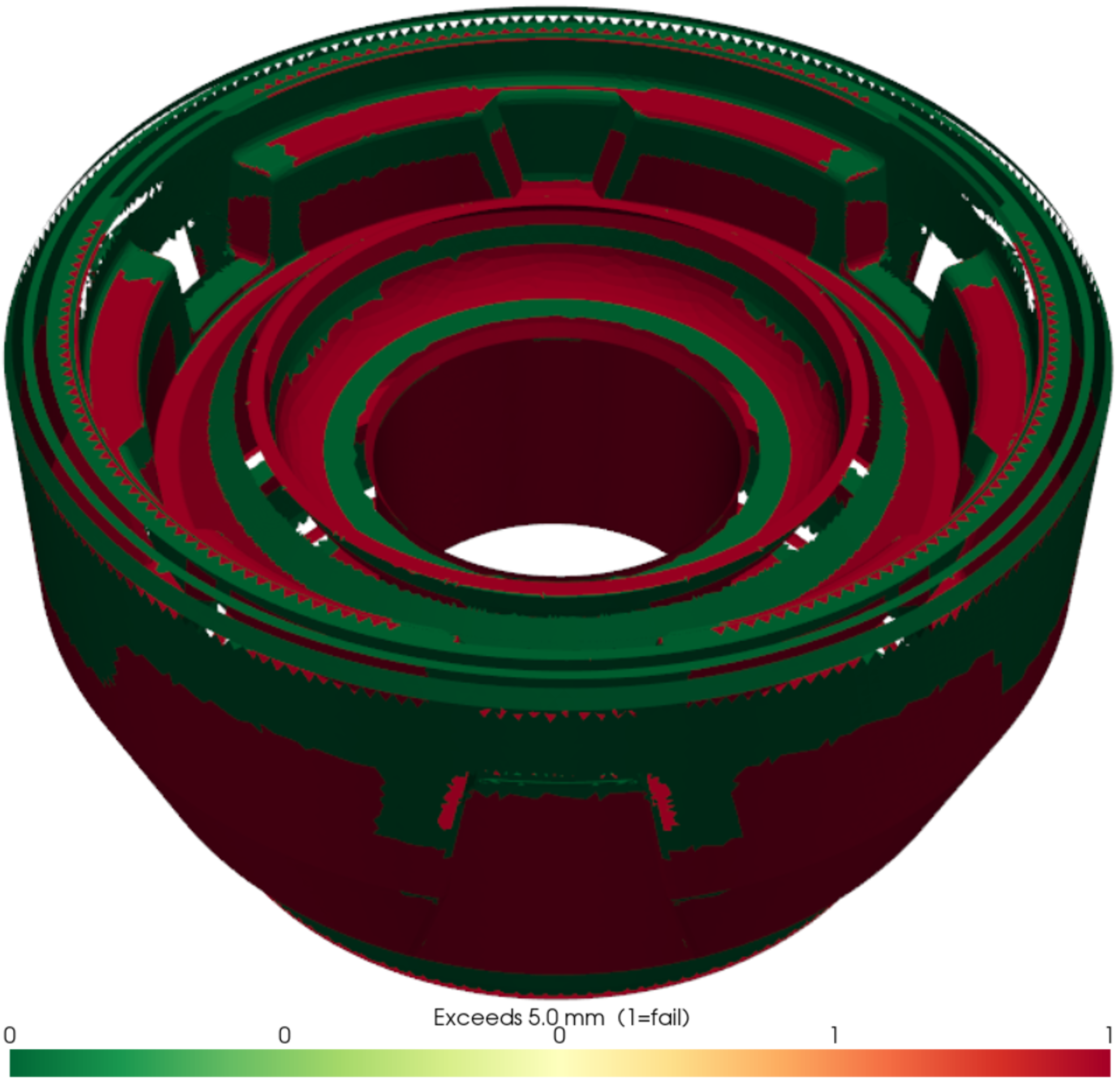}  
		\caption{Rule 1: Maximum Thickness}
		\label{fig:maxthickness}  
	\end{figure}

	\begin{figure}[h]  
		\centering
		\includegraphics[width=0.35\textwidth]{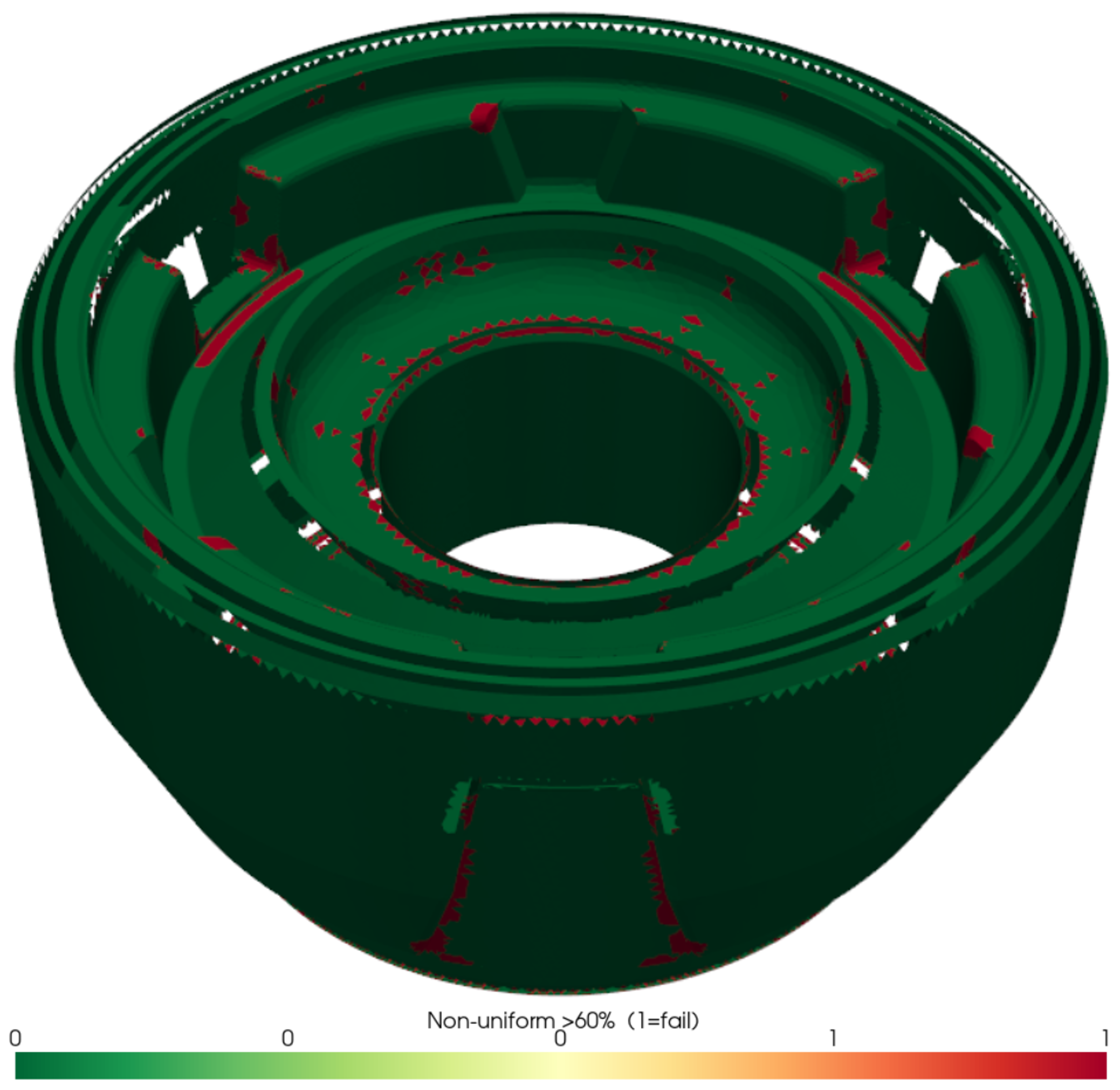}  
		\caption{Rule 2: Uniformity}
		\label{fig:uniformity}  
	\end{figure}
	
	\begin{figure}[h]  
		\centering
		\includegraphics[width=0.35\textwidth]{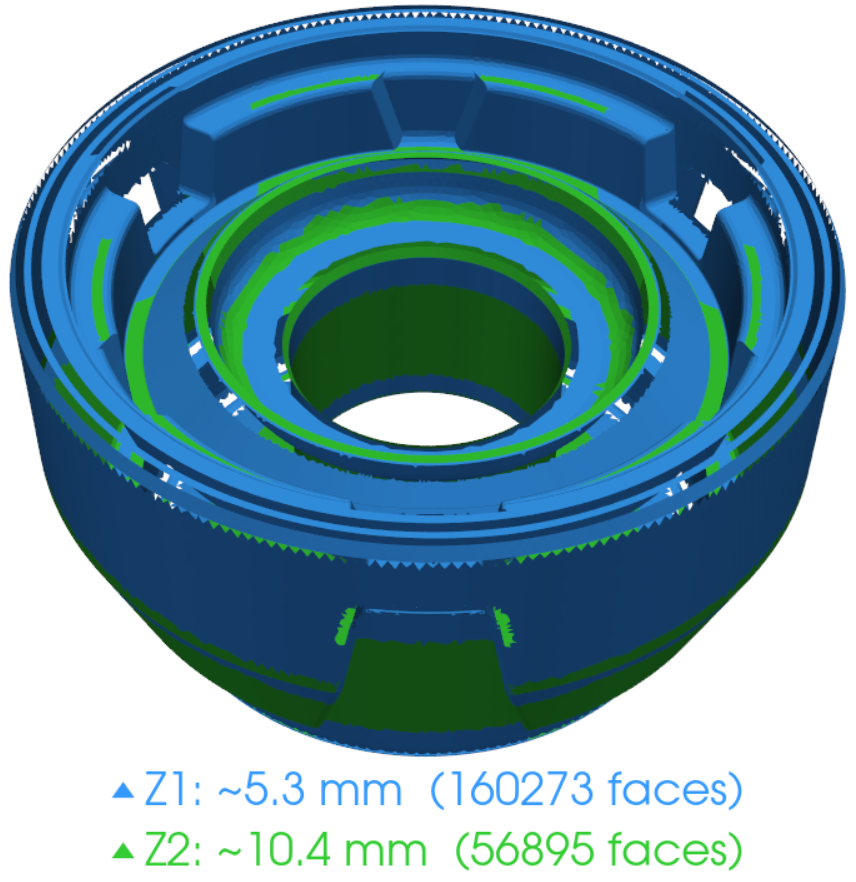}  
		\caption{Thickness Zones}
		\label{fig:zones}  
	\end{figure}

	Rule evaluation results are summarized in Table~\ref{tab:moldingrules} 
	and shown in Figures~\ref{fig:maxthickness} and~\ref{fig:uniformity}. The 
	maximum thickness threshold is set at 5~mm, and the uniformity rule at 
	60\%, as indicated, which is an unrealistic scenario but was selected to 
	depict the analysis' capabilities.
	Thickness zone clustering produces two significant zones, shown in 
	Figure~\ref{fig:zones}. For each thickness zone, a similar statistical 
	summary can be extracted, as the one shown in Table~\ref{tab:moldingstats}.

	\subsection{Milling Case Study}
\begin{figure}[t!]  
		\centering
		\includegraphics[width=0.44\textwidth]{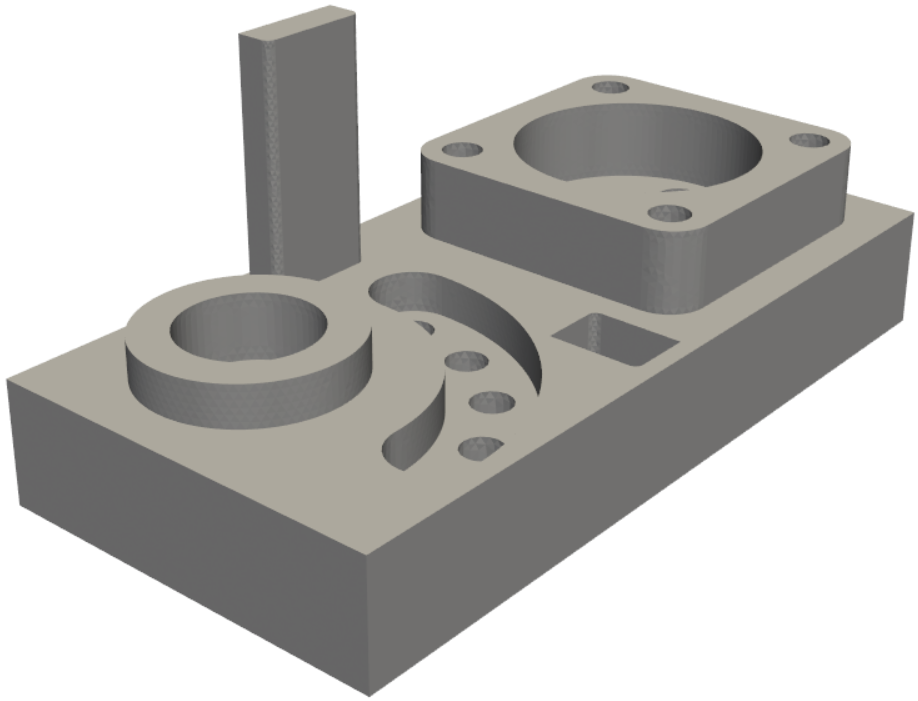}  
		\caption{Milling - Case Study Part}
		\label{fig:millingpart}  
	\end{figure}

\begin{figure}[h]  
		\centering
		\includegraphics[width=0.44\textwidth]{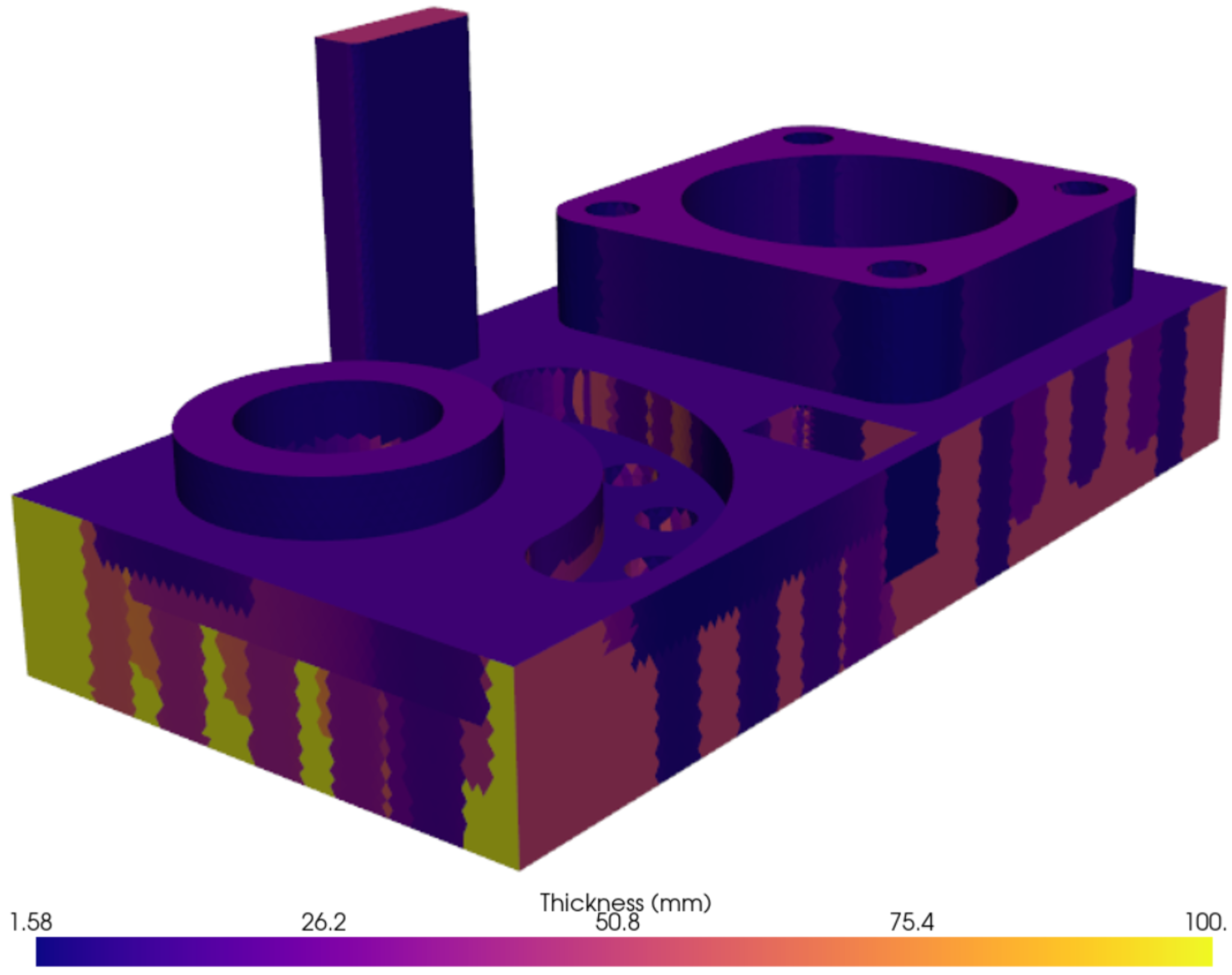}  
		\caption{Ray-based Thickness Heatmap}
		\label{fig:millingheatmap}  
	\end{figure}

	The milling case study part is shown in Figure~\ref{fig:millingpart}. 
	The analysis is performed on a STEP file imported, tessellated with 
	a target of 50,000 triangles, using the thresholds listed in 
	Table~\ref{tab:millingparams}. The resulting ray-based thickness 
	heatmap is shown in Figure~\ref{fig:millingheatmap}.

	\begin{table}[b]
		\centering
		\begin{tabular}{l c}
			\hline
			\textbf{Parameter} & \textbf{Value} \\
			\hline
			Absolute thickness -- warning level & 5.0~mm\\
			Absolute thickness -- critical level & 1.5~mm\\
			Height-to-thickness ratio -- warning level & 3.0\\
			Height-to-thickness ratio -- critical level & 20.0\\
			\hline
		\end{tabular}
		\caption{Analysis parameters for the milling case study.}
		\label{tab:millingparams}
	\end{table}
	
	\begin{figure}[h!]  
		\centering
		\includegraphics[width=0.44\textwidth]{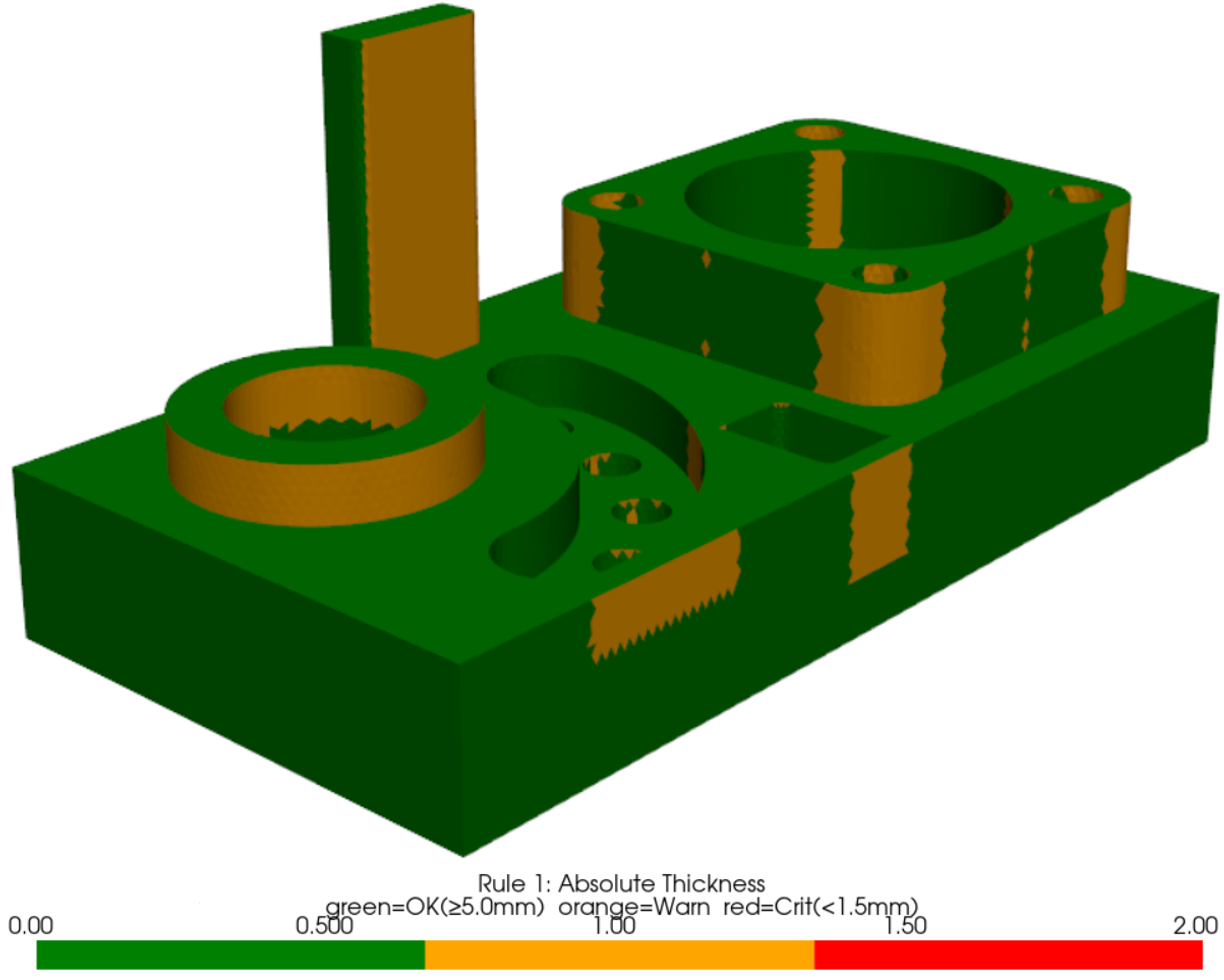}  
		\caption{Rule 1: Minimum Thickness}
		\label{fig:millingmin}  
	\end{figure}
	
	\begin{figure}[h!]  
		\centering
		\includegraphics[width=0.44\textwidth]{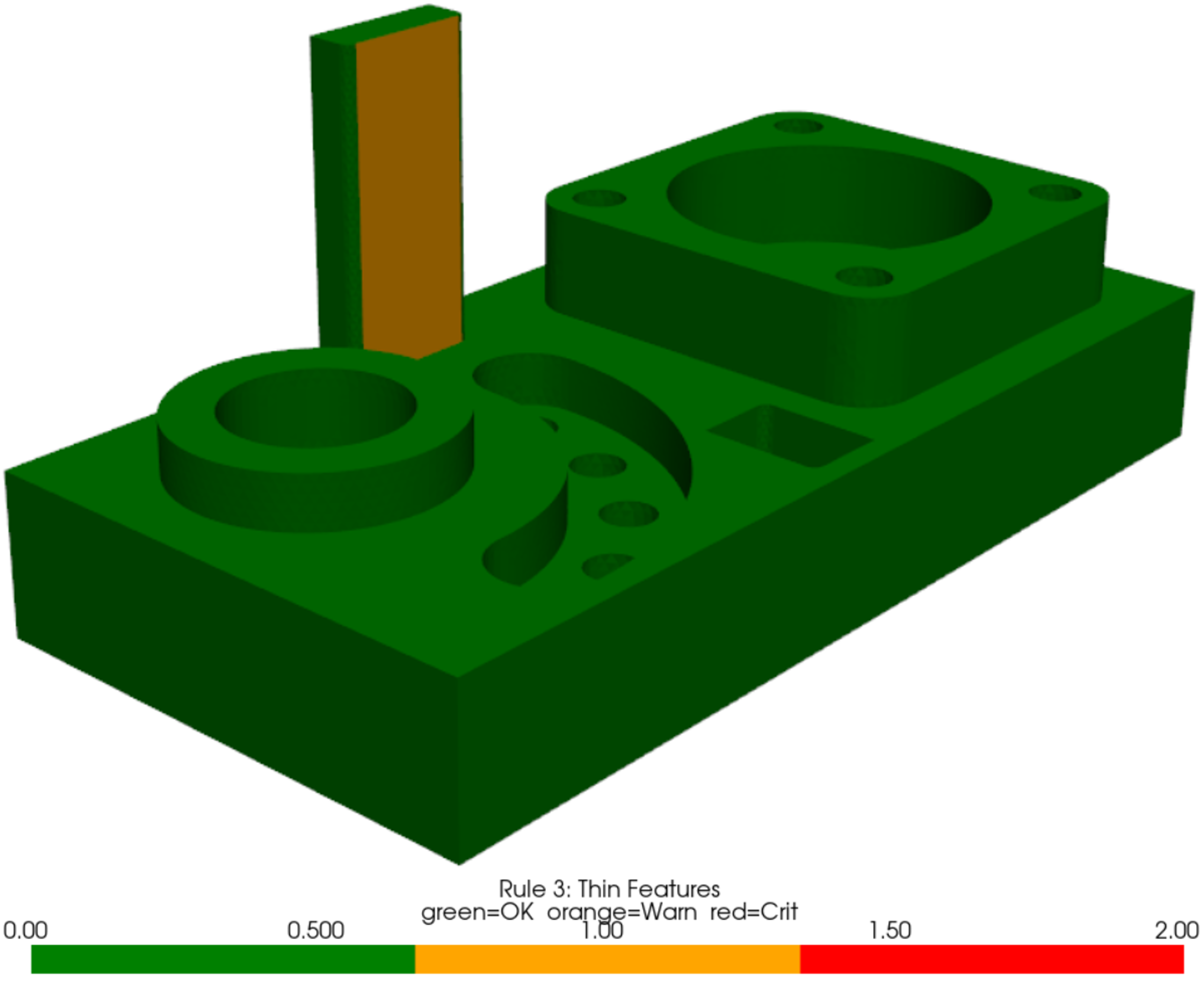}  
		\caption{Rule 2: Thin Feature Detection}
		\label{fig:thinfeatures}  
	\end{figure}
	
	\begin{table}[h]
		\centering
		\begin{tabular}{l c}
			\hline
			\textbf{Metric} & \textbf{Value} \\
			\hline
			Minimum thickness & 1.58~mm\\
			Maximum thickness & 100.00~mm\\
			Mean thickness & 18.53~mm\\
			Standard deviation & 3.34~mm\\
			\hline
		\end{tabular}
		\caption{Thickness statistics for the milling case study.}
		\label{tab:millingstats}
	\end{table}
	
	The summary statistics are reported in Table~\ref{tab:millingstats}, and 
	processed further under DFM milling criteria.
	Rule~1 evaluates minimum thickness across all matched triangles. As 
	shown in Figure~\ref{fig:millingmin}, the part shows no critical error, but 
	several warning facets. Rule~2 detects parallel face pairs constituting 
	thin features. For this step, mapping between BRep faces and tessellated 
	facets is required, in order to have an accurate evaluation of the 
	height-to-thickness ratio for each surface. As shown in 
	Figure~\ref{fig:thinfeatures}, 1 thin feature is identified.

	\section{Discussion}
	
	The two case studies demonstrate the system's ability to surface 
	process-specific thickness violations on real industrial parts, 
	producing per-face diagnostic results and zone-level summaries that 
	would be directly actionable in a DFM workflow.
	
	The molding analysis identifies a part with significant thickness 
	accumulation, concentrated in two distinct thickness zones. The wall uniformity 
	rule flags 5.9\% of facets, indicating that thickness transitions 
	are largely gradual across the part, with only localised abrupt 
	changes. These results, taken together, suggest a part where 
	excessive wall thickness is the primary manufacturability concern.
	
	The parameters used in this case study are deliberately set to 
	expose a large fraction of violations, demonstrating the system's 
	sensitivity across a wide range of part geometry. In practice, a 
	designer would calibrate these thresholds to material-specific 
	guidelines and would use the 
	zone-level statistics to target redesign efforts on the most 
	problematic regions.
	
	A key contribution of the molding branch is the automatic boundary 
	face filtering, which in this case removes the 6.2\% of facets that the sphere method systematically underestimates 
	thickness \cite{inui2016, inui2024}. This filtering is performed 
	without user intervention by deriving the minimum threshold from 
	the mean sphere-based thickness of facets adjacent to boundary convex edges, making the 
	analysis robust to this known failure mode. However, the approach 
	assumes that boundary facets are representative of the underestimation 
	magnitude, which may not hold for all part geometries. Parts with 
	very irregular bounding box extents may require manual calibration of the edge factor and margin 
	parameters to avoid over- or under-filtering. This remains a 
	limitation of the current implementation.
	
	A further limitation concerns the uniformity rule. The rule operates on all valid facets without spatial connectivity 
	filtering, meaning that small isolated facets with thickness 
	values diverging from their neighbours may be flagged as non-uniform. 
	This behavior is intentional, as the uniformity rule is designed to 
	surface all abrupt local thickness changes for designer review, 
	regardless of whether they belong to a coherent zone. The designer 
	is expected to inspect flagged facets and determine whether a redesign 
	is warranted, rather than relying on automated pass/fail 
	classification alone.
	
	The milling analysis identifies a part with no triangles falling below 
	the critical threshold of 1.5~mm. The 17.4\% of triangles classified 
	as warning under Rule~1 indicate regions where wall thickness 
	approaches but does not breach the critical limit, representing 
	areas of potential concern under aggressive cutting conditions. 
	Rule~2 detects one thin feature from parallel face pair analysis. 
	
	The 100.0~mm maximum thickness 
	value reported in the summary statistics is a known artefact of the 
	ray-based method, as in regions where no directly opposing surface 
	exists along the inward normal, the ray travels without intersection 
	until a distant surface is reached, returning a spuriously large 
	value \cite{sinha2007}. As discussed in the methodology, this artefact 
	does not affect rule evaluation, since both DFM rules target minimum 
	thickness and feature slenderness rather than maximum values.
	
	The thresholds used in this case study are once again
	selected to demonstrate the system's classification granularity 
	across warning and critical severity levels. In a production 
	setting, these would be derived from material properties, tool 
	diameter constraints, and machine rigidity specifications, and 
	could be adjusted per feature type or machining setup.
	
	A key contribution of the milling branch is the combination of 
	triangle-level absolute thickness assessment with BRep-level thin 
	feature detection. The ray-based thickness provides a dense, 
	spatially continuous diagnostic map of the part surface, while the 
	parallel face pair analysis adds a geometric reasoning layer that 
	identifies structurally vulnerable features not necessarily apparent 
	from thickness alone. A tall, slender rib may pass the absolute 
	thickness check while remaining susceptible to deflection under 
	cutting forces. The height-to-thickness ratio check surfaces this 
	risk explicitly.
	
	For the completion of the aforementioned analysis, a current limitation of the milling branch is its dependency on the 
	STEP file format for thin feature detection. The BRep face extraction 
	and face-to-triangle mapping required by Rule~2 rely on the 
	BRep surface information present in STEP files, which is not 
	available in STL format. A user working with an STL-only workflow 
	can still obtain Rule~1 thickness results, but thin 
	feature detection is unavailable without the corresponding STEP 
	file.
	
	The explicit process selection at the entry point of the 
	system ensures that the diagnostic output is always interpreted 
	against the correct failure criteria for the intended manufacturing 
	process. The system has been implemented as an interactive tool, 
	publicly available at \cite{github}, accepting standard CAD file 
	formats with user-adjustable parameters and GPU-accelerated 
	computation, making it suitable for integration into existing DFM 
	workflows without requiring specialist simulation software or 
	extended computation times. Installation instructions and usage 
	documentation are provided in the repository accordingly.

	\section{Conclusion and Future Work}
	
	This paper presented a process-aware thickness analysis system for CAD 
	parts intended for molding and milling, in which the geometric estimation 
	method and the DFM rule set are jointly determined by the target 
	manufacturing process. A process-driven method selection strategy was proposed and implemented, 
	in which the sphere-based method is applied to molding parts to detect 
	maximum thickness violations, wall non-uniformity, and thickness zone 
	structure, while the ray-based method is applied to milling parts to 
	assess minimum thickness and identify thin features through 
	parallel face pair detection. Each method is applied where its geometric 
	strengths align with the DFM criteria of the target process, and its 
	known failure modes fall outside the region of diagnostic interest. Validation on two real industrial parts 
	demonstrated the system's ability to surface process-relevant 
	manufacturability issues with configurable severity classification.
	
	Several directions are identified for future work. First, automatic 
	threshold derivation would replace user-defined DFM thresholds with 
	values retrieved from a material and process database, reducing the 
	configuration burden on the designer and improving consistency across 
	analyses. Second, filtering of spurious maximum thickness values in 
	the milling branch, arising from rays that find no directly opposing 
	surface, would improve the completeness of the thickness statistics 
	reported, complementing the existing minimum thickness rule evaluation. 
	Third, integration with CAD environments would allow 
	diagnostic results to be linked directly to the feature tree of the 
	originating model, enabling violation-driven redesign suggestions 
	rather than surface-level flagging alone. These extensions would 
	further close the gap between geometric thickness analysis and 
	actionable DFM feedback within the design cycle.

	\FloatBarrier

\end{document}